\documentclass[aps,prd,preprint,groupedaddress,showpacs]{revtex4}
\usepackage{epsfig}
\usepackage{latexsym}

\begin{document}

\title{Relativistic Hubble's law}
\author{Yongkyu Ko}
\email[]{yongkyu@phya.yonsei.ac.kr}
\homepage[]{http://phya.yonsei.ac.kr/~yongkyu}
\affiliation{
Department of Physics, Yonsei University, Seoul 120-749, Korea}
\date{\today}

\begin{abstract}
Considering the hyperbolic nature of the universe, the Hubble's law and the
inverse square laws such as, the Coulomb's law and Newton's gravitational law,
should be modified in accordance with the special theory of relativity.
Consequently there is no the Hubble's length, which might be the observable
boundary of our universe, does not appear in point of view
of the special theory of relativity, and the Newton's third law still
hold in the special theory of relativity.  Recent astronomical observations of
type Ia supernovae support the modified Hubble's law which also leads a view
that the cosmic microwave might come from extra-universes.
\end{abstract}

\maketitle

Recent astronomical observations of type Ia supernovae out to a redshift of
$z \leq 1$
\cite{Riess,Garnavich,Perlmutter,Perlmutter2,Perlmutter3} show that the
receding velocities of the supernovae are observed to be not linear to their
distances.  Since the supernovae are observed to be fainter than the
luminosities at the expected distances by the linear Hubble's law, they are
assumed to be farther than the distances.  It is interpreted that the universe
is accelerating or inflating under the assumption of the flat universe
\cite{Riess,Perlmutter,Perlmutter2}.  The observational results
may well be interpreted as the inflation, but there is a serious doubt: Stars in
the universe are neither rockets nor missiles, then what is the repulsive force
between stars?  The cosmological parameter $\Omega_\Lambda$ once used by
Einstein for the static universe is insisted to play the role of the force as
the vacuum energy density, the zero point fluctuations of quantum fields,
the potential energy density of dynamic fields, or a network of cosmic strings
\cite{Starkman}.
According to the theory, our universe is still on the way of the Big Bang and
not expanding after the Big Bang.

If our universe is not flat, the story becomes a completely different story.
The observations of type Ia supernovae may play a key role in determining the
geometry of the universe.  Since type Ia supernovae are used as the standard
candles in astronomy, they are good distance indicators in the universe by
measuring the luminosity of them \cite{Sandage}.  Furthermore
special relativity tells us the geometry of the universe which was early
recognized by H. Minkowski and others \cite{Einstein1905,Sexl}.
However the series of the follower's works on the hyperbolic geometry are
not paid attention to by many physicists \cite{Walter}.  The recent
work of the author shows that the geometry of the universe is hyperbolic quite
simply by investigating the geometric aspect of the Lorentz transformation.
A Lorentz transformation in momentum space can be obtained from hyperbolic
trigonometric relations by simply multiplying a mass to the relations \cite{Ko}.
These trigonometric relations are for the triangle made by an observed moving
particle and the two observing inertial frames which are moving relative to
each other.  This triangle is a hyperbolic triangle
so that the sum of its interior angles is less than $\pi$.  It should be noted
that the interior angles are not measured in one inertial frame, but done in
the  each inertial frame at the three vertices of the triangle.
Since the constantly moving particle can be thought to be in its inertial frame,
the hyperbolic triangle means that the universe to which the three inertial frames
belong can be regarded as a hyperbolic space in the limit of no gravitation.

The hyperbolic nature of the universe is also shown toward the limit of no
matter density in the Freidmann-Robertson-Walker cosmology which comes from the
general theory of relativity.  Since the gravitational force is attractive,
it is expected that the rate of velocity to distance of nearby stars should be
less than that of distant stars.  On the contrary to the expectation, the
observations show that the rate of velocity to distance of nearby stars larger
than that of distant stars \cite{Riess}.  It can be thought that the
gravitation between stars nearly play any role in the expansion of the universe.
Therefore the geometry caused by the matter of the universe can be ignored to
a first approximation.  The recent observations of type Ia supernovae
are explained well in view of hyperbolic space time.

\begin{figure}
\centerline{\epsfig{file=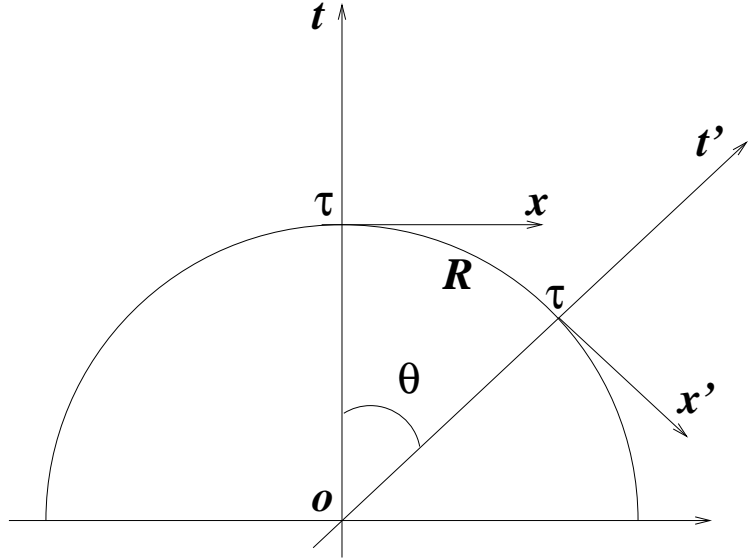, width=0.6\columnwidth}}
\caption{This is a schematic diagram for hyperbolic space and two inertial
frames, which are moving relative to each other.  The origins of the
two inertial frame was coincided at the moment $t=t'=\tau=0$.  The rest frame
is represented with the coordinate $(t,x)$, and the moving frame whose velocity
is $v = \tanh \vartheta = \tanh R/\tau$ is done with $(t',x')$.  The other
coordinates $y$ and $z$ are supposed to have the same relation with the time
axis.}
\label{fig1}
\end{figure}

Since a hyperbolic space is the space observed simultaneously in the invariant
time, which might be similar to the conformal time \cite{Bucher}, by using the
Lorentz transformation, a hyperbolic space can be depicted
schematically on a hemisphere as shown in Fig. \ref{fig1} at a certain instant.
The space time diagram, of course, does not exactly reflect reality,
because there is no way to draw a
diagram exactly in hyperbolic space even in the Poincar\'{e} model and the
Klein model.  Only part of the complete hyperbolic nature can be drawn.
The diagram only tells us that the space time is intrinsically curved, the two
inertial frames have the same invariant time and so on, but does not say the
complete nature of hyperbolic space.
The Lorentz transformation should be written in terms of correct vectors in each
inertial frame \cite{Ko} by
\begin{eqnarray}
& &t' = t \cosh \vartheta + \hat{n} \cdot \mbox{\boldmath{$r$}}
\sinh \vartheta, \nonumber \\
& &\hat{n}' \cdot \mbox{\boldmath{$r$}}' = t \sinh \vartheta
+ \hat{n} \cdot \mbox{\boldmath{$r$}} \cosh \vartheta, \nonumber \\
& &\hat{n}' \times (\hat{n}' \times \mbox{\boldmath{$r$}}')
= \hat{n} \times (\hat{n} \times \mbox{\boldmath{$r$}}), \label{usualvectr}
\end{eqnarray}
where $\hat{n}$ and $\hat{n}'$ are the directions of the relative velocities
observed in the two inertial frames for the origin of each other frame,
respectively.
The magnitudes of the relative velocities are the same as $v = \tanh \theta$ in
both inertial frames.

From this Lorentz transformation, the sphere in Fig. 1 satisfies the following
relation:
\begin{equation}
t'^2 -(\hat{n}' \cdot \mbox{\boldmath{$r$}}')^2
- \{ \hat{n}' \times (\hat{n}' \times \mbox{\boldmath{$r$}}') \}^2
= t^2 -(\hat{n} \cdot \mbox{\boldmath{$r$}})^2
- \{ \hat{n} \times (\hat{n} \times \mbox{\boldmath{$r$}}) \}^2 =\tau^2,
\end{equation}
which can be simplified as
\begin{equation}
t'^2-r'^2 = t^2 - r^2 = \tau^2,
\end{equation}
where
$\tau$ means the invariant time which have elapsed since the instant coincided by
the two frames.  If the direction of the relative velocity is taken to be $x-$axis,
the above equation is more familiar form as follows
\begin{equation}
t^2-x^2  = t'^2-x'^2  =\tau^2.
\end{equation}
If the space axis is considered to be imaginary axis, then the above equation
is nothing but an equation of a circle or a sphere.  Here the metric plays the
role of an imaginary axis like $\hat{n} \cdot \hat{n}$.  After the time $\tau$
have elapsed, two inertial frames can be described on the hyperbolic sphere as
the coordinate systems depicted in Fig. 1.

Now what is the distance between the origins of the two frames?
$x=vt$, $x'=vt'$ or something else.  According to the time dilation or the
length contraction, the first two distances are not equal to each other.
The observed distances can be different from the real distance in a curved space,
because only the origin of the coordinate is coincided with the real space.
The definition for a distance between two points which are in motion relatively
should be independent of coordinate systems.  The lack of uniqueness for the
definition of the distance between the two points may have a serious problem.
As an example an electron, which is coming to the earth very fast in the sun,
sees a proton in the rest frame where we live, but does not interact with it,
even though the distance in the moving frame of the electron is approximately
zero due to the length contraction and vice versa.  Alternatively it can be
thought that the electron interacts with the proton very slowly due to the time
dilation.  A result of length contraction can attribute to a result of time
dilation. However the problem is not solved but handed over the problem of
time.  If then, as another example, the age of the universe needs not to be
found out, because it, {\it a priori}, varies from inertial
frame to inertial frame, namely, form zero to infinity.

There exists a distance to be able to be reconciled between the two frames in
the text of hyperbolic geometry\cite{Henle,Coxeter}.
A distance between two points in the hyperbolic plane of an unit circle in an
inertial frame is defined as
\begin{equation}
d={1\over2} \ln {|1-z_1z_2|+|z_1-z_2|
\over{|1-z_1z_2|-|z_1-z_2|}},\label{hyper}
\end{equation}
where $z_1$ and $z_2$ is in the complex plane which is mapped from the real
hyperbolic plane.  Therefore in the above situation,
if $z_1=0$, $z_2=x/t=v$, and $d=R/\tau$ are inserted into Eq. (\ref{hyper}),
then the relation $R/\tau = 1/2 \ln(1+v)/(1-v) = \tanh^{-1}v$ can be obtained.
The inverse to this relation is $v = \tanh (R/\tau)$.  The distance between
the two frames is defined as $R$ and satisfies the following equation:
\begin{equation}
v = {x \over t} = {x' \over{t'}} = \tanh{R \over{\tau}}=\tanh \theta,
\label{hdistance}
\end{equation}
and the real distance is defined to be identical in the two frames
whether we think it in the rest or moving frames.
Therefore the fast moving electron in the sun need not dance with the
fluctuating proton in the earth and vice versa, if the Coulomb's law together
with the Newton's gravitational law is understood as the
inverse square law of the real distance.  Since the distance between the
origins of two frame is the same, the Newton's third law still holds in
relativistic kinematics.

This phenomenon might be examined in the accelerators such as SLAC \cite{European}.
The accelerator is a linear type so that the frame of an accelerated particle is
a good moving frame and a target is in a good rest frame.
Electrons in the accelerator can obtain 50 GeV energies.  Hence the
velocities of the electrons are accelerated up to 0.99999999995 times light velocity.
The length contraction in the moving frame is 0.00001 times the length observed
in the rest frame.  However it sounds strange that there is a report that the
accelerated particles are scattered off before arriving at the target due to
the length contraction.

The Hubble's law tells us that stars in the universe are observed to be moving
away from the observing position, whose velocity is proportional to their
distance \cite{Freedman}.  The data of Ref. \cite{Freedman} shows that the
velocity is fitted to be proportional to the distance up to 450 Mpc.
This is a record that our universe have been expanding with a constant rate
of velocity to distance since the Big Bang.
This Hubble's law can be regarded as a
non-relativistic one, because a star can move away with the more rapid
speed than the speed of light over a certain distance.
The limit to be able to observe the universe is known as the
Hubble length $L_0=c/H_0$ \cite{Harrison} or the minimal antitrapped
surface(MAS) \cite{Starkman}.  Recent observations do not show
the linearity but show the curve of Eq. (\ref{hdistance}), because the motion
of the stars in the region of a redshift $0.1<z<1$ is sufficiently relativistic.

In another aspect the linear Hubble's law is not universal all over the
universe,
but provides us with different values for the Hubble's constant according to
inertial frames.  Using the velocity addition rule this fact can be understood
with ease.  If we observed a star $s_1$ with a velocity $v$ and a
distance $d$, an observer in the star $s_1$ would be expected to observe
another star $s_2$ with the same
velocity and distance from the same Hubble's law.  So this star $s_2$ is
observed that the velocity is $2v/(1+v^2)$ and the distance is
$d+d/\sqrt{1-v^2}$ in our inertial frame.  However, according to the linear
Hubble's law,
the distance of the star  $s_2$ is calculated as $2d/(1+v^2)$ for the velocity
$2v/(1+v^2)$.  This is quite different from the relativistic result.  If we
observe the n-th star like the above way, what is the velocity and distance of
the star?  It is an interesting finite series as follows
\begin{eqnarray}
v_n &=& {v_{n-1}+v \over{1+v_{n-1}v_1}}=\tanh n \theta, \nonumber \\
d_n &=& d(1+{1 \over{\sqrt{1-v_1^2}}}+ \cdot \cdot \cdot+
{1 \over{\sqrt{1-v_{n-1}^2}}})
={d \over2}{e^{n \theta}-e^{(1-n) \theta}+e^{\theta}-1 \over{e^{\theta}-1}},
\end{eqnarray}
where the calculations can be done by taking the initial velocity
as $v=v_1=\tanh \theta$.  If the initial velocity and distance are sufficiently
small, then we can calculate approximately
the ratio $v/d \approx \theta/d = H_0$ and the relation of the n-th star
as $v_n = \tanh H_0 d_n$.
Therefore the linear Hubble's law can not be compatible with the
special theory of relativity.

If the velocities of stars have not been changed since the Big Bang, then the
ratio $R/\tau$'s of the stars are constant and the same with $dR/d\tau$'s,
respectively.  Naturally the Hubble's law should be modified as
\begin{equation}
v =\tanh \theta = \tanh{R \over{\tau}} = \tanh{H_0 R},
\end{equation}
where  $H_0$ is the Hubble constant and $R$ is the distance of a star
from the observing position as shown in Fig. 1.  This modified Hubble's law
gives a constant curvature $c \tau=c/H_0$ in any spatial direction to
the universe.  This modified law do not have the observable boundary of the
universe, that is, the universe is infinite.
Therefore several cosmological phenomena could be interpreted in
a little different way, for
examples, darkness at night, cosmic microwave background and so on.

If the universe is infinite, infinite numbers of stars would be shining the
earth like day.  Therefore night is not dark as it is.  There is a proof done by
Einstein in the footnote of the sentence: The stellar universe ought to be a
finite island in the infinite ocean of space \cite{Einstein}.  In the proof the
inverse square law of the luminosity only holds in flat universe.  The law
should be modified in hyperbolic space as
\begin{equation}
{\cal F} = {{\cal L} \over{ 4 \pi \tau^2 \sinh^2 {R \over{\tau}}}}
= {{\cal L} \over{ 4 \pi {\sinh^2 H_0 R \over{H_0^2}}}}.
\end{equation}
where ${\cal L}$ is the luminosity of the standard candle and ${\cal F}$ is the
observed flux of the source.
The luminosity of the stars at a distance is fainter than in the case of
the inverse square law.  However this can not be the reason of the darkness at
night, because the gauss law in hyperbolic space also gives constant outgoing
flux with respect to distance.  The suitable reason might be the redshift of
receding stars.  The looking back time \cite{Harrison} is not suitable reason
for darkness at night, because the universe is infinite and because there was
other universes before the Big Bang.
If the light of such stars are extremely
redshifted in accordance with the Hubble's law, then the light would
be observed as extremely longer electromagnetic waves.  The
extreme case can be regarded as an alternative interpretation of the source of
the cosmic microwave background.  The highly
redshifted light of usual stars could be the source of the cosmic microwave
background.

The redshift is told as $z=1100$ \cite{Bond2}.
According to the linear Hubble's law, the distance of the source of cosmic
microwave background would be around 4996.54 Mpc from the earth for a Hubble
constant $H_0 = 60$
km/sMpc.  If the stars in the universe is distributed uniformly, it is
questionable that the source of cosmic microwave is enough to fill the universe
with cosmic microwave in the shell of around the radius of $4996.54$ Mpc.
The modified Hubble's law gives even more thick shell of the source of cosmic
microwave for the same redshift range, because the distance of the same redshift
amounts to 34995.8 Mpc, if the Hubble constant is the same over all universes.
Since the distance is even farther than the $c/H_0 = 4996.54$ Mpc, the cosmic
microwave might come from the source in extra-universes and be the signal
occurred 114 billion years ago before the Big Bang,
because $\tau = 1/H_0=16.3$ billion years is the time when our universe
is in a point according to the Big Bang theory.  The COBE maps \cite{Bennett}
and the map of the
angular correlation patterns on CMB anisotropy \cite{Bond,Bond2} give this
interpretation an interesting thought that the shape might be the images of
the extra-universes, instead of the question \cite{Cornish}: Can COBE see
the shape of the universe?  If there were a birth of another universe
through a Big Bang, the signal might be the longer wavelength than CMB.

The luminosity-distance law can be also confirmed from the observation of the
number of stars or galaxies within some solid angle of a telescope with respect
to the distance.
If the universe is homogeneous and isotropic so that the number of stars are
uniformly distributed, then the number of stars within a distance $R$ would be
proportional to $H_0^{-3} \sinh^3 H_0R$.

Besides the relativistic Hubble's law and the luminosity-distance law, there
are some possible corrections due to the relativistic motions of supernovae.
Among them, it is difficult to ignore the correction due to the time dilation
which, is shown in the light curve fit for a $1+z$ time dilation \cite{Riess},
causes the luminosity of supernovae fainter.  Since the luminosity is a
radiating power of a supernova, the radiation energy is independent of
the receding velocity of a supernova.  If  a supernova with a redshift $z$ is
observed, the luminosity is related with the luminosity of a supernova at rest
as follows
\begin{equation}
dE= {\cal L}_0 dT_0 = {\cal L}dT,
\end{equation}
where $E$ is the radiation energy and $T_0$ and $T$ means the time of the rest
frame and the moving frame, respectively.  Hence the luminosity of a moving
supernova would be observed as follows
\begin{equation}
{\cal L} = \sqrt{1-v^2} {\cal L}_0={2 (z+1) \over{(z+1)^2+1}}{\cal L}_0.
\label{timedi}
\end{equation}
This correction gives 80 \% observed luminosity for a supernova of a redshift
$z=1$.
There are also many astronomical corrections for the moving supernova which is
called K-correction.  The K-correction has a dependence with a redshift of a
supernova
due to the fact that the relative photon fluxes of high-redshift supernovae are $1+z$
``brighter" than energy fluxes \cite{Nugent} and the K-correction  is a purely
technical effect \cite{Oke}.  Since the redshift dependence of Eq. (\ref{timedi}) in
the luminosity is different from that of the K-correction, the above correction
must be certainly not included in the K-correction.

Another important correction is due to the
redshift of the spectral distribution for a supernova.  If we know the spectral
distribution of type Ia supernovae or at lest their temperature, we can correct
the luminosity by using the spectral distribution of supernova or the Plank
distribution, respectively.
As an example an observed spectrum of wavelength $4000-7000$ \AA~ for a redshift
$z=1$ comes from the lights of wavelength $2000-3500$ \AA~ in the spectrum of the
supernova which is quite different from the visible light in magnitude in the
spectral distribution of the supernova.  The type Ia supernovae can not be
standard candles without the knowledge of the spectral distribution of the
supernovae, because the luminosity-distance law is reliable to the same
luminosity of a observed wavelength width.  If we observed a spectrum of a
wavelength $\lambda$ through a filter from a supernova which has a redshift
$z$, then the luminosity of the supernova should be corrected with the ratio:
\begin{equation}
R(z) = u(\lambda/(1+z))/u(\lambda), \label{corr2}
\end{equation}
where $u(\lambda)$ is the Plank distribution function.  Actually all the light
curves of high-redshift type Ia supernovae \cite{Riess} show that the light
curve of V
photometry is brighter than that of B photometry.  This means that the observed
spectrum comes from the shorter wavelength side than the wavelength of its
maximum in the spectral distribution.

\begin{figure}
\centerline{\epsfig{file=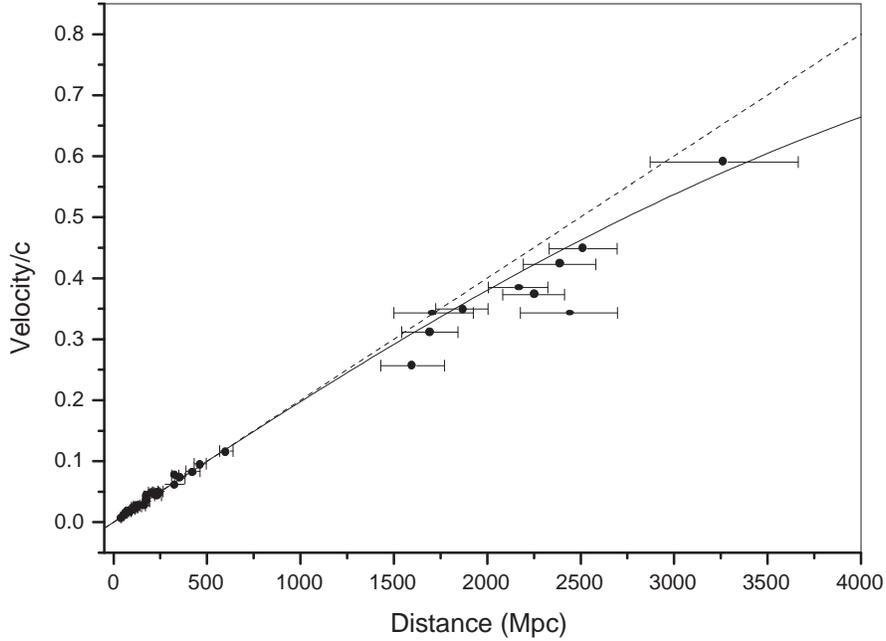, width=0.8\columnwidth}}
\caption{These are the graphs for the receding velocity to the distances for the
recent observed supernovae.  The data are taken from Ref. \cite{Riess} and
corrected in accordance with Eqs. (\ref{timedi}) and (\ref{corr2}).  The dashed
line is a linear curve and the solid line is a hyperbolic tangent curve for
a Hubble constant 60 km/sMpc.  In the correction the maximum wavelength 4200
\AA~
of B photometry and the temperature 10000 ${}^0$ K of a Plank distribution function
are used. }
\label{fig2}
\end{figure}

Fig. \ref{fig2}, where the distances are calculated from the distance moduli
``$\mu_0$" of Ref. \cite{Riess} and applied to the above corrections, shows that
the above interpretation explains the recent observations of type Ia
supernovae \cite{Riess} well.  It would be more reliable, if the real spectral
distribution of supernovae is used.
There are still many reasons coursing distant stars fainter than the expected
brightness in the universe.  The opacity of the vacuum of the universe, which is
not considered here, might be another reason of the faintness of stars.
The data in the figure do not still shows clear distinction between the linear
and the hyperbolic curves, but future searches for more redshifted supernovae
can confirm the modified Hubble's law.  Those work also can search the
evidence of deceleration of the universe due to the gravitational effect.
The general theory of relativity or the Freidmann-Robertson-Walker cosmology
can play an important role in the problem of expanding universe after such works.

The establishment of the Hubble's law is a first step in astronomy,
astrophysics and cosmology for giving reasonable physical view to our
understanding on the universe.  If the law is confirmed certainly, the redshift
would be a good distance indicator.  Moreover this law may change our current
view on the universe.  Since the motion of supernovae in the region of redshift
$0.1 < z < 1$ is relativistic, it is necessary to correct the Hubble diagram
with the relativistic effects as shown above.  Among these corrections the
Hubble diagram is very sensitive to the correction using a spectral distribution
function.  Therefore it is inevitable to know the exact spectral distributions of
nearby type Ia supernovae and necessary to make a template for this correction
like the template of light curve \cite{Riess}.

The interpretation that our universe is accelerating is very premature.  The
data show that the Hubble diagrams can be fitted to be linear for a redshift
$z<0.1$ and for a redshift $0.1<z<1$, respectively, without the above
corrections and has different Hubble constants(see Fig. 2 in Ref.
\cite{Perlmutter2}), if the diagrams are drawn separately in the two regions.
This means that our universe
was accelerated around 1.6 billion years ago at the time when the stars of a
redshift $z=0.1$ sent their lights to us.
It is necessary to explain the reason why the universe should be accelerated
especially at that time in such an interpretation or to show the relic that
the vacuum energy density was inflated or accelerated at the special time.

\end{document}